\UseRawInputEncoding
\documentclass[aps,prb,superscriptaddress,twocolumn]{revtex4-2}

\usepackage{graphicx}
\usepackage{amsfonts,amsmath,amssymb,amsthm}
\usepackage{epstopdf}
\usepackage{upgreek,xspace}
\usepackage{chngcntr}
\usepackage{hyperref}
\hypersetup{colorlinks,allcolors=blue}
\usepackage[version=3]{mhchem} 
\usepackage{soul}

\usepackage{setspace}

\newcommand{\be}{\begin{equation}}
\newcommand{\ee}{\end{equation}}

\newcommand{\Tc}{$T_\mathrm{c}$}
\newcommand{\Hc}{$H_\mathrm{c2,\perp}$}
\newcommand{\Tg}{$T_\mathrm{G}$}
\newcommand{\Po}{$P_\mathrm{O_2}$}
\newcommand{\AO}{Al$_2$O$_3$}

\newcommand{\dC}{$^\circ$C}
\newcommand{\MAO}{MgAl$_2$O$_4$}
\newcommand{\aO}{$a_\mathrm{O}$}
\newcommand{\MNO}{Mg$_4$Nb$_2$O$_9$}

\begin{document}

\title{Superconducting vacancy-ordered rock-salt NbO films}

\author{Jeong~Rae~Kim}
\affiliation{Department of Applied Physics and Materials Science, California Institute of Technology, Pasadena, California 91125, USA.}
\affiliation{Institute for Quantum Information and Matter, California Institute of Technology, Pasadena, California 91125, USA.}

\author{Sandra~Glotzer}
\affiliation{Department of Applied Physics and Materials Science, California Institute of Technology, Pasadena, California 91125, USA.}

\author{Evan~Krysko}
\affiliation{Department of Materials Science and Engineering, Cornell University, Ithaca, New York 14853, USA.}
\affiliation{Platform for the Accelerated Realization, Analysis, and Discovery of Interface Materials (PARADIM), Cornell University,
Ithaca, New York 14853, USA}

\author{Matthew~R.~Barone}
\affiliation{Department of Materials Science and Engineering, Cornell University, Ithaca, New York 14853, USA.}
\affiliation{Platform for the Accelerated Realization, Analysis, and Discovery of Interface Materials (PARADIM), Cornell University,
Ithaca, New York 14853, USA}

\author{Jinkwon~Kim}
\affiliation{Department of Materials Science and Engineering, Cornell University, Ithaca, New York 14853, USA.}

\author{Salva~Salmani-Rezaie}
\affiliation{Department of Materials Science and Engineering, The Ohio State University, Columbus, Ohio 43210, USA}

\author{Adrian~Llanos}
\affiliation{Department of Applied Physics and Materials Science, California Institute of Technology, Pasadena, California 91125, USA.}

\author{Joseph~Falson}
\email{falson@caltech.edu}
\affiliation{Department of Applied Physics and Materials Science, California Institute of Technology, Pasadena, California 91125, USA.}
\affiliation{Institute for Quantum Information and Matter, California Institute of Technology, Pasadena, California 91125, USA.}


\begin{abstract}
We report molecular beam epitaxy synthesis of vacancy-ordered rocksalt NbO thin films which display superconductivity. A comparative study of substrates identifies \AO (0001) as the optimal platform for realizing high-quality, single-phase films when growing at temperatures exceeding 1000~$^\circ$C. The controlled NbO films exhibit superconductivity with critical temperatures up to \Tc=1.37~K, comparable to bulk single crystals. This work addresses the fundamental bottlenecks encountered in the high-temperature epitaxy of compounds with uncommon oxidation states, while expanding the scope of available thin-film superconductors. 
\end{abstract}

\flushbottom
\maketitle

\section*{Introduction}
The tuning of crystalline systems towards uncommon oxidation states presents a novel synthetic pathway to control electronic phases in condensed matter science~\cite{khomskii2014transition,imadaMetalinsulatorTransitions1998b}. As the oxidation state is directly linked to the charge-transfer energy, control over the orbital occupancy can exert a substantial influence on the resulting electronic structure and band filling~\cite{zaanen1985band}. The inherent instability associated with uncommon charge configurations can additionally place compounds in the vicinity of exotic electronic phase transitions which are tunable with temperature or pressure~\cite{cox1976crystal,johnston2014charge,sun2023signatures}. Certain rare crystal structures are uniquely stabilized by these uncommon oxidation states~\cite{wang2020antiperovskites}, some of which have been studied extensively as thin films where the substrate finds utility as a stabilizing template, and a range of reducing or oxidizing conditions can be achieved when leveraging distinct aspects of the growth process~\cite{li2019superconductivity,kim2023geometric,ko2024signatures,mairoser2015high,jeen2013reversible}.

The focus of this work is the transition-metal monoxide NbO, which is special for numerous reasons~\cite{hulm1972superconductivity,okaz1975specific,efimenko2017electronic}. It is the sole well-characterized compound having the Nb$^{2+}$ oxidation state, while simultaneously being the only well-characterized compound crystallizing in the stoichiometric NbO-type structure~\cite{burdett1984niobium,liu2014first,zhou2022synthesis}. In contrast to the insulator Nb$_2$O$_5$ and metal-insulator transition hosting NbO$_2$~\cite{janninck1966electrical}, the 4$d^3$ filling in NbO results in a highly metallic system with bulk crystals displaying high residual resistivity ratios exceeding 100 and a low-temperature resistivity on the order of 0.1~$\mu \Omega \mathrm{.cm}$. As shown in Fig.~\ref{Fig1}, the crystal is the so-called vacancy-ordered rock-salt structure in which a quarter of the ions in the regular rock-salt lattice are removed. Combined with its superconductivity below 1.61~K, NbO emerges as an attractive material platform for studying the interplay between its uncommon oxidation state, unique crystal structure, and superconductivity. And while the structure favors formation as a stoichiometric compound with NbO$_{1\pm \delta}$ ($\delta\approx0.02$) \cite{okaz1975specific}, tuning growth conditions conducive with stabilizing the Nb$^{2+}$ oxidation state poses challenges in material synthesis. Indeed, NbO synthesis can be accompanied by the formation of elemental Nb and NbO$_2$ defect phases~\cite{hulm1972superconductivity,valeeva2023niobium}, which complicate the interpretation of the electronic properties of the material, including the superconducting transition temperature.

Here, we report the epitaxy of single-phase vacancy-ordered rocksalt NbO thin films which display superconductivity. In line with bulk synthesis, precise control of oxidation states at high temperatures enables highly crystalline NbO thin films. We emphasize that proper selection of the substrate material and growth plane is essential to enable systematic oxidation control of the Nb-O system. Superconducting characteristics comparable to bulk single crystals are achieved in (111)-oriented epitaxial NbO films grown on \AO~(0001) substrates.

\section*{Results and discussion}
\subsection*{Overview of growth}
\begin{figure}[t]
  \centering
   \includegraphics[width=85mm]{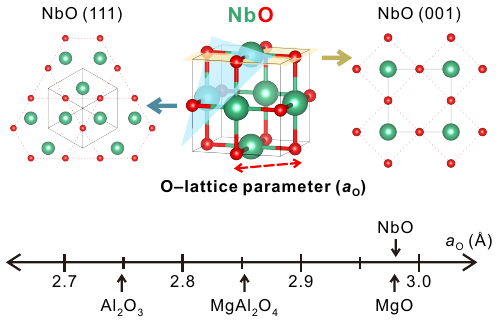}
   \caption{Crystal structure of vacancy-ordered rock-salt NbO and two film growth planes; the (001) plane forms a square lattice while the (111) plane forms a pseudo-triangular lattice. The nearest-neighbor O-O spacing being the O-lattice parameter ($a_\mathrm{O}$) is compared between NbO and MgO,~\MAO, and~\AO~substrates.}
   \label{Fig1}
\end{figure}

Given the unconventional Nb$^{2+}$ oxidation state in NbO, it is reasonable to expect that reducing growth conditions will be required for stabilizing single-phase films. To that end, we have employed molecular beam epitaxy (MBE), working in a growth regime of high substrate temperatures (growth temperature, \Tg) and low molecular oxygen partial pressures. The former is achieved using laser heating, while the latter is quantified using the local pressure proximate to an injection nozzle (nozzle pressure, \Po), in a similar configuration to our previous study \cite{kim2025high}. Nb metal flux is provided using electron beam evaporation. The next consideration is the crystallographic relationship between NbO and a template substrate, which in this study are MgO, \MAO, and \AO. These have cubic rocksalt, cubic spinel, and hexagonal corundum structures, respectively. We also consider NbO (001) and (111) as growth planes. In order to compare these structurally distinct materials on the same line, we define an oxygen lattice parameter (\aO) as shown in Fig.~\ref{Fig1}, which allows us to consider these different scenarios as  square or triangle lattices of oxygen ions. Figure \ref{Fig1} shows the square and pseudo-triangle oxygen lattices of NbO (001) and (111). The \aO~could be either strictly calculated or averaged to be 2.977 \AA~(NbO and MgO), 2.856 \AA~(\MAO), and 2.748 \AA~(\AO). In these configurations, NbO would be subjected to $-$0 \% (MgO), $-$4.1 \% (\MAO), and $-$7.7 \% (\AO) compressive epitaxial strains. 

\begin{figure*}[t]
  \centering
   \includegraphics[width=170mm]{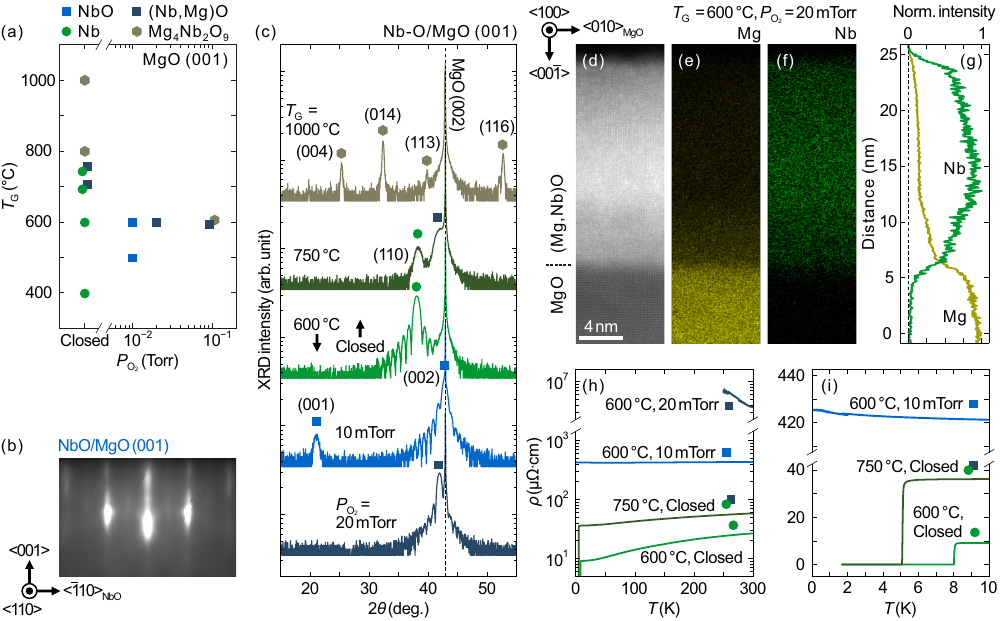}
   \caption{Growth of Nb-O on MgO (001). (a) A growth phase diagram in the~\Po-\Tg~parameter space with symbols corresponding to single-phase NbO (blue square), rock-salt like (Nb,Mg)O alloy (dark blue square), body-centered-cubic Nb (green circle), and Mg$_4$Nb$_2$O$_9$ (khaki hexagon). (b) RHEED image of a single-phase NbO sample measured along the MgO [110] direction. (c) Out-of-plane XRD features of representative films: Mg$_4$Nb$_2$O$_9$ [(\Tg,~\Po) = (1000~\dC, `Closed')], mix-phase Nb-(Nb,Mg)O [(\Tg,~\Po) = (750~\dC, `Closed')], Nb [(\Tg,~\Po) = (600~\dC, `Closed')], NbO [(\Tg,~\Po) = (600~\dC, 10~mTorr)], and (Nb,Mg)O [(\Tg,~\Po) = (600~\dC, 20~mTorr)]. HAADF-STEM image (d) of the (Nb,Mg)O/MgO (001) and EDS mappings of Mg (e) and Nb (f) ions along with their normalized values (g), taken from the MgO [100] zone axis. (h,i) Temperature-dependent resistivity measurements of the samples in (c). The Mg$_4$Nb$_2$O$_9$/MgO (001) was insulating. 
   }
   \label{Fig2}
\end{figure*}

\subsection*{MgO (001)}
Conventional epitaxy approaches would suggest rock-salt MgO is the canonical choice of the substrate for NbO films due to the structural relevance, near 0 \% strain mismatch, and thermal stability. Moreover, many transition-metal monoxides have successfully been grown on MgO (001) substrates~\cite{gao1998preparation,rata2004growth,budde2018structural,li2021single}. Here, however, we demonstrate it is poorly suited for NbO, as summarized in Fig.~\ref{Fig2}. Four different phases are apparent in our range of \Po~and \Tg~variations presented in the phase diagram in Fig.~\ref{Fig2}a. While working with the nozzle closed, Nb films form below \Tg~= 600 \dC, but Mg-Nb-O alloys formed at higher \Tg. The coexistence of metallic Nb and a Mg-Nb-O alloy is resolved in the intermediate \Tg~= 600--800 \dC~range based on X-ray diffraction (XRD) fingerprints at 2$\theta$~=~41.5 $^\circ$~in 2$\theta$-$\omega$ scans (Fig.~\ref{Fig2}c). This phase is denoted as (Nb,Mg)O throughout this work. Above \Tg~= 800~\dC, a polycrystalline \MNO~phase additionally emerges whose XRD peaks could be assigned to those of previous reports (Fig.~\ref{Fig2}c)~\cite{ananta2004synthesis}. The trend clearly points to significant interdiffusion between substrate and film, similar to previous works operating at high growth temperatures in other crystal systems~\cite{gao1998preparation,budde2018structural,kim2025high}.

While high \Tg~evidently drives alloying, working at lower \Tg~= 600 \dC~and with finite \Po~is found to be a pathway towards single-phase NbO stabilization. Using (\Tg, \Po) = (600 \dC, 10 mTorr), singly (001)-oriented NbO emerges and displays the expected peak positions (Fig.~\ref{Fig2}c). The Laue oscillations and the reflection high-energy electron diffraction (RHEED) pattern indicate good uniformity and a two-dimensional surface (Figs.~\ref{Fig2}b and c). Higher oxygen supply further is seen to induce alloying, with a single-phase (Nb,Mg)O forming at (\Tg, \Po) = (600 \dC, 20 mTorr) and a mixed (Nb,Mg)O-\MNO~phase forming at (\Tg, \Po) = (600 \dC, 100 mTorr).

For further insight into the (Nb,Mg)O phase, we turn to cross-section scanning transmission electron microscopy (STEM) for the (\Tg, \Po) = (600 \dC, 20 mTorr) sample. The high-angle annular dark-field STEM (HAADF-STEM, Fig.~\ref{Fig2}d) and energy dispersive spectroscopy (EDS, Figs.~\ref{Fig2}e-g) were performed along MgO [100] zone axis. Although a thorough structural identification requires additional investigation, the cross-sectional feature of the (Nb,Mg)O phase appears analogous to the rock-salt structure of MgO substrate layer. The STEM-EDS analysis shows a low but persistent Mg signal throughout the film layer. As two possible explanations for the (Nb,Mg)O phase, we suggest a NbO-MgO rock-salt alloy and a Mg$_3$Nb$_6$O$_{11}$ phase. Recently, stabilization of a rock-salt NbO film was reported~\cite{kimura2024elevating}. A transition metal monoxide, NiO, is known to form continuous rock-salt solid solutions with MgO~\cite{prostakova2012experimental}. Taking them together, we hypothesize that NbO-MgO rock-salt solid solutions may form during the growth process. Reports of the Mg$_3$Nb$_6$O$_{11}$ phase (Nb oxidation state = 2.67+) are limited within the literature~\cite{burnus1987mn3nb6o11}. Although the structure is hexagonal, we noted that a (1$\overline{1}$1)-oriented growth of Mg$_3$Nb$_6$O$_{11}$ can mimic the XRD and STEM characteristics to some extent [(2$\overline{2}$2) peak 2$\theta$ = 42.1 \d~and (011) as a STEM cross section]. Considering the interfacial region with higher Mg concentration (Fig.~\ref{Fig2}g), the coexistence of both is also plausible.

Temperature-dependent resistivity measurements of representative samples are shown in Figs.~\ref{Fig2}h and i. The Nb/MgO (001) [(\Tg, \Po) = (600 \dC, `Closed')] qualitatively reproduces the known high conductivity and superconducting transition temperature (\Tc~$\approx$~9.3 K) of elemental Nb. Contrary to expectations, the NbO/MgO (001) [(\Tg, \Po) = (600 \dC, 10 mTorr)] shows more than two orders of magnitude higher resistivity than bulk NbO values and was not superconducting down to 0.1 K~\cite{hulm1972superconductivity,honig1973electrical}. The (Nb,Mg)O/MgO (001) [(\Tg, \Po) = (600 \dC, 20 mTorr)] and \MNO/MgO (001) [(\Tg, \Po) = (1000 \dC, `Closed')] were insulating. The resistivity and superconductivity (\Tc~$\approx$~5 K) of the Nb-(Nb,Mg)O/MgO (001) [(\Tg, \Po) = (750 \dC, `Closed')] can likely be interpreted as Nb metal affected by the (Nb,Mg)O defects.

The absence of superconductivity in the NbO/MgO (001) is surprising and could perhaps be overcome by further growth optimization. However, the Mg-Nb-O alloying in high \Tg~or high \Po~conditions restricts extended searches of the parameter space. Therefore, despite its excellent structural match, we disqualify MgO as a substrate for superconducting NbO films.    

\begin{figure*}[t]
  \centering
   \includegraphics[width=170mm]{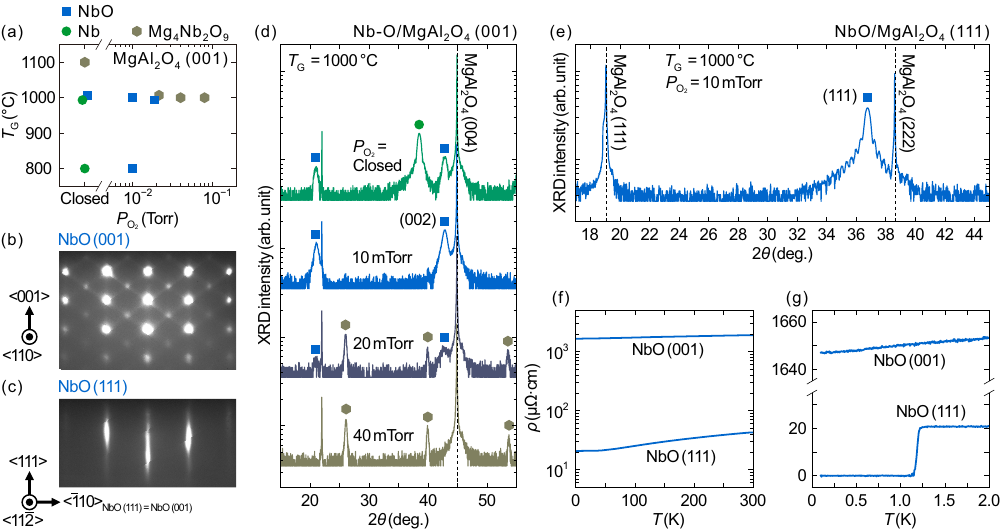}
   \caption{Growth of Nb-O on~\MAO~(001) and (111). (a) Growth phase diagram in the~\Po-\Tg~parameter space. (b,c) RHEED images of single-phase NbO/\MAO~(001) (b) and (111) (c) samples measured along~\MAO~[110] (b) and ~\MAO~[11$\overline{2}$] (c) directions. (d) Out-of-plane XRD of the Nb-O/\MAO~(001) samples grown at~\Tg~=~1000~\dC~and~\Po~= `Closed', 10~mTorr, 20~mTorr, and 40~mTorr. (e) Out-of-plane XRD of the NbO/\MAO~(111) grown at ~\Tg~=~1000~\dC~and~\Po~=~10~mTorr. (f,g) Temperature-dependent resistivity of NbO/\MAO~(001) and (111).
   }
   \label{Fig3}
\end{figure*}

\begin{figure*}[t]
  \centering
   \includegraphics[width=170mm]{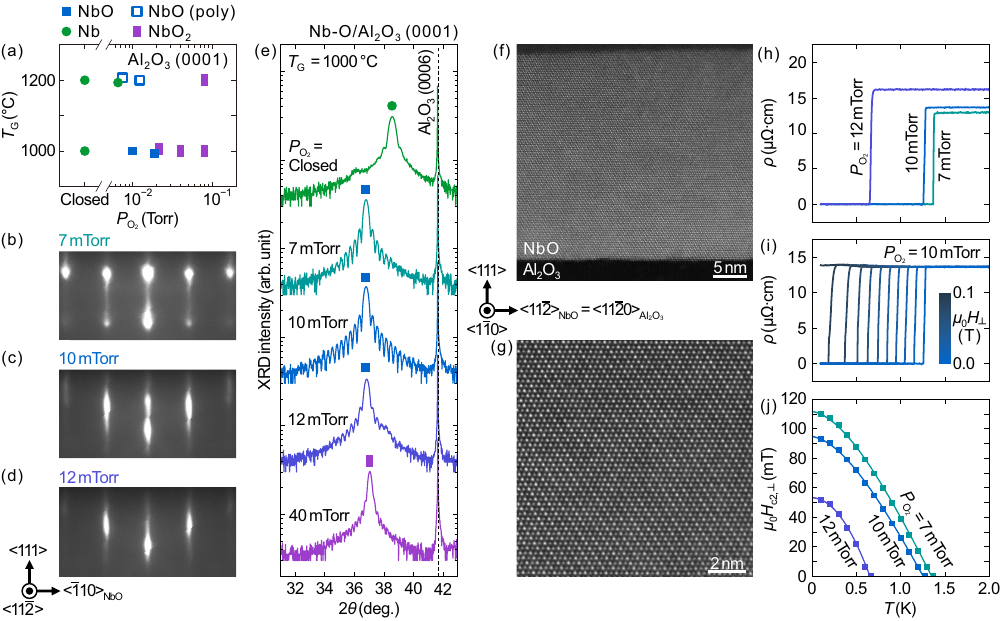}
   \caption{Growth of Nb-O on~\AO~(0001). (a) Growth phase diagram in the~\Po-\Tg~parameter space. (b,c,d) RHEED images of samples grown at~\Tg~=~1000~\dC~and~\Po~=~7~mTorr (b), 10~mTorr (c), and 12~mTorr (d), measured along the~\AO~[11$\overline{2}$0] direction. (e) Out-of-plane XRD of the Nb-O/\AO~(0001) samples grown at~\Tg~=~1000~\dC~and~\Po~= `Closed', 7~mTorr, 10~mTorr, 12~mTorr, and 40~mTorr. (f,g) HAADF-STEM images of the 10~mTorr sample in (e) taken from the~\AO~[1$\overline{1}$00] zone axis. (h) Temperature-dependent resistivity measurements of samples in (e). (i) Temperature-dependent resistivity of the 10~mTorr sample under an out-of-plane magnetic field. (j) Out of plane critical field versus temperature of the 7, 10, and 12~mTorr samples. Solid lines are Werthamer-Helfand-Hohenberg (7 and 10~mTorr) and Gorter-Casimir fits (12~mTorr)~\cite{werthamer1966temperature,gorter1934supraconductivity}.
   }
   \label{Fig4}
\end{figure*}

\subsection*{\MAO~(001) and (111)}
We now consider \MAO~as a candidate substrate in Fig.~\ref{Fig3}, while extending the parameter space to include the (001) and (111) facets. The spinel \MAO~has a cubic lattice parameter $a$~= 8.078 \AA, which can facilitate rocksalt monoxide growth through an epitaxial relationship with half of the \MAO~lattice~\cite{rata2005strain}. Performing a similar investigation of the \Po-\Tg~parameter space (Fig.~\ref{Fig3}a) using \MAO~(001), Mg-Nb-O alloy formation is relatively suppressed with the use of \MAO~substrates. The phase diagram is qualitatively similar to that of MgO (001), but the threshold for alloy formation is shifted to approximately 1000 \dC. In the \Po~= `Closed' sample series, a Nb metal phase is observed at low \Tg~= 800 \dC, whereas a mixed Nb-NbO phase is observed at \Tg~= 1000 \dC~(Fig.~\ref{Fig3}d). This can be explained by thermally activated oxygen diffusion from the \MAO~substrate~\cite{kim2025high}. Single-phase NbO/\MAO~(001) is achieved at \Po~= 10 mTorr, consistent with the growth on MgO (001). Although we did not observe the (Nb,Mg)O phase in this experimental range, the \MNO~phase was seen to form at high \Tg~or high \Po~conditions.  

The high-temperature growth using \MAO~substrates provides new insights into the growth dynamics of NbO films. Unlike NbO/MgO (001), the growth of NbO/\MAO~(001) was three-dimensional as evidenced by the RHEED pattern (Fig.~\ref{Fig3}b) showing clear transmission pattern of the cubic lattice. The broad XRD peaks [NbO (001) and (002)] also suggest marginally low crystallinity. Although surface energy calculations for NbO have not been reported to our knowledge, the charge neutrality of the NbO (001) surface is clearly broken by the ordered vacancies. Thus, while the (001) facet of a full rock-salt structure is lower in surface energy~\cite{wander2003stability,liu2006surface}, this is not guaranteed in the case of NbO. This hypothesis is supported by a recent single-crystal study that demonstrates that (111) is the natural NbO growth plane~\cite{valeeva2023niobium}. To explore this hypothesis, results on \MAO~(111) are presented in Figs.~\ref{Fig3}c and e. Remarkably, the NbO/\MAO~(111) exhibits a streaky two-dimensional RHEED pattern and sharp NbO (111) XRD peak accompanied by well-defined Laue oscillations and no additional diffraction peaks.

The temperature-dependent resistivity of the NbO/\MAO~(001) and (111) films are compared in Figs.~\ref{Fig3}f and g, confirming the critical role of the growth plane. The resistivity of the NbO/\MAO~(001) is about two orders of magnitude higher than that of the NbO/\MAO~(111). More notably, while the NbO/\MAO~(001) is not superconducting down to 0.1~K, the NbO/\MAO~(111) manifests a superconducting transition at 1.18~K, defined as the temperature at half-resistance. Given that substrate orientation was the sole variable, we infer that the pronounced degradation of the electrical properties arises from the structural irregularities associated with the high-energy NbO (001) surface.

The NbO growths on \MAO~substrates unveil that the chemical stability and film morphology can be greatly improved depending on the substrate material and orientation, even at the cost of an increased lattice mismatch. The clear selectivity toward the lower energy surface is a sign of thermodynamically driven enhancement of the NbO crystallinity. However, the low-temperature resistivity and \Tc~of the superconducting NbO/\MAO~(111) is still inferior to the bulk values (0.1 $\mu\Omega\cdot$cm and 1.61 K)~\cite{hulm1972superconductivity,honig1973electrical}. The proximity to alloy formation in the parameter space may imply that trace amounts of Mg-O diffusion are possible even in the superconducting NbO/\MAO~(111) sample.

\begin{table*}[t]
\centering
\caption{
	\label{table} 
	Superconductivity parameters of NbO (111) samples. All samples are 30$\pm$1 nm thick.
      }
        \begin{tabular}{c|ccc}\hline\hline
          Samples &  ~~~~\Tc(0) (K)~~~~ & ~~~~\Hc(0) (T)~~~~ & ~~~~$\xi$(0) (nm)~~~~\\\hline
          NbO/\MAO~(111) (\Po~= 10 mTorr) & 1.18 & 0.141 & 48.3\\
          NbO/\AO~(0001) (\Po~= 7 mTorr) & 1.37  & 0.112 & 54.2\\
          NbO/\AO~(0001) (\Po~= 10 mTorr) &  1.27 & 0.095 & 58.9\\
          NbO/\AO~(0001) (\Po~= 12 mTorr) &  0.67 & 0.053 & 78.7\\
    \hline\hline
  \end{tabular}
\end{table*}

\subsection*{\AO~(0001)}
A common thread between the use of MgO and \MAO~has been the alloy formation between Mg and NbO, in various configurations. It is notable that in Fig.~\ref{Fig3} there was no evidence of alloying with Al-O species. Therefore, we hypothesize that despite the largest lattice mismatch of approximately $-$7.7 \%, \AO~may be a suitable candidate for growth. This is bolstered by evidence suggesting that (111) growth leads to superior crystallinity, which will naturally be promoted on the hexagonal (0001) facet of \AO. Indeed, we note that NbO (111) growth on \AO~(0001) substrates has been reported to a limited extent in a previous study of the Li-Nb-O system~\cite{tellekamp2017molecular}.

The Nb-O/\AO~(0001) growths are summarized in Fig.~\ref{Fig4}. This phase diagram is distinguished from the former two in that it reflects only the Nb-O system without any alloys (Fig.~\ref{Fig4}a). In contrast to MgO and \MAO~where excessive oxidation led to alloy formation, here we can see the formation of a rutile NbO$_2$ phase in high-\Po~conditions (Fig.~\ref{Fig4}e)~\cite{bolzan1994powder}. While the NbO$_2$ (200) peak appear very similar to NbO (111) in 2$\theta$-$\omega$ scans, asymmetric diffraction can help differentiate between the phases (Fig.S2 and Table 2). In this system the growth is evidently limited by a new factor, that is the formation of polycrystalline samples at \Tg~= 1200 \dC, which could be attributed to the large strain mismatch of $-$7.7~\%. 

Fine tuning of \Po~at \Tg~=1000 \dC~gives us access to the highest quality (111) films. Here we focus on three samples, grown with \Po~= 7, 10, and 12 mTorr, as shown in Figs.~\ref{Fig4}b--d, and consider surface versus bulk crystallinity. The \Po~= 12 mTorr sample displays the most well-defined Ewald's sphere while the RHEED pattern of the \Po~= 7 mTorr sample shows mostly transmissive spots. These features coexist in the RHEED pattern of the \Po~= 10 mTorr sample, which is also the sample to show the most pronounced Laue oscillations and symmetric peak shape in 2$\theta$-$\omega$. Meanwhile, the Laue oscillations observed in the \Po~= 7 and 12 mTorr samples were relatively suppressed, and the \Po~= 12 mTorr sample displays a subtle additional feature on the high-2$\theta$ side of the NbO (111) peak. These data suggest subtle sample-to-sample variations in crystal quality; XRD suggests the \Po~= 10 mTorr sample is structurally optimal, with imperfections in the \Po~= 7 and 12 mTorr samples presumably related to excess Nb and O ions. Variations in RHEED however are likely associated with surface-localized off-stiochiometries; the three-dimensional surface features are attributed to excess Nb (Fig.~\ref{Fig4}b) while the O-rich sample is likely to have an optimal surface composition. In Fig.~\ref{Fig4}f and g we present HAADF-STEM on the (\Tg, \Po) = (1000 \dC, 10 mTorr) sample along \AO~[1$\overline{1}$00] zone axis. A sharp NbO/\AO~heterointerface aligns with the lack of alloying discussed above. Importantly, the vacancy order rocksalt structure is clearly resolved in the magnified view (Fig.~\ref{Fig4}g).

All three of the single-phase NbO/\AO~samples display superconductivity with sharp transitions. Figures \ref{Fig4}h--j present temperature- and field-dependent resistivity measurements for the samples with the superconducting \Tc~varying between 1.37 (\Po~= 7 mTorr), 1.27 (\Po~= 10 mTorr), and 0.62 K (\Po~= 12 mTorr). The out-of-plane critical field (\Hc) scales with \Po~values in a similar fashion (Fig.~\ref{Fig4}j). The zero-temperature \Hc(0) and coherence length [$\xi$(0)] values are summarized in Table 1. Pinpointing the origin of the \Tc~increase in the \Po~= 7 mTorr sample will be the subject of a future detailed investigation into transport properties. The Nb-rich growth conditions can yield either oxygen-deficient NbO$_{1-\delta}$ or Nb-NbO mixed phases. Compared to \Tc~= 1.61 K for NbO$_{1.00}$, NbO$_{0.96}$ shows suppressed superconductivity with \Tc~= 1.37 K~\cite{okaz1975specific}, nearly matching our highest \Tc~for the Nb-rich \Po~= 7 mTorr sample. Beyond the solubility limit of Nb in NbO, the \Tc~enhancement is markedly stronger~\cite{hulm1972superconductivity}.

\subsection*{Discussion}
Despite \AO~(0001) being the least structurally compatible with NbO qualitatively (cubic/hexagonal) and quantitatively ($-$7.7 \% strain mismatch), it permits exploration of the largest thermodynamic parameter space through enhanced chemical stability and suppressed alloy formation. This supports the realization of the highest quality films achieved in this study. However, these crystals are achieved in narrow MBE growth windows with the divergence of optimal structure, optimal surface, and optimal superconductivity samples indicating the lack of a self-limited growth process similar to those exploited to control such defect-sensitive superconductors~\cite{nair2018demystifying,ahadi2019enhancing}. Ideally, identifying a  diffusion-controlled growth process similar to our previous report in the Ti-O system~\cite{kim2025high} could achieve this, and indeed glimpses of this process are evident in the \MAO~(001) (Figs.~\ref{Fig3}a,d), but full realization is impeded by alloy and polycrystal formations at higher temperatures. Additionally, twin domain formation and the large lattice mismatch are two major structural imperfections that remain. Twin domains are observed in all NbO (111) samples including the NbO/\MAO~(111), an iso-symmetrical cubic/cubic heterostructure (Fig. S3). Given the dearth of commercially available alternatives, we expect the development of suitable epitaxial buffer layers could address many of these outstanding issues and deliver significant enhancements in crystal quality and an expanded growth window. The ideal buffer layer is required to have a cubic crystal structure, small lattice mismatch, while being chemically stable at the interface with the host substrate and subsequent NbO layer up to very high temperatures~\cite{hubbard1996thermodynamic}. This is likely to be an increasingly important problem to address as new materials come into focus which benefit from high synthesis temperatures and reducing synthesis conditions~\cite{kim2025high}.

\section*{Conclusion}
The distinctive chemical and structural characteristics of NbO have presented several challenges in its growth as thin films. We have approached synthesis of this compound by suppressing film/substrate alloying and surface roughening issues, thereby identifying \AO~as the leading substrate for its high-temperature synthesis. The vacancy-ordered rock-salt structure in thin films is unambiguously demonstrated, as is superconductivity. NbO stands out as a rare example of a stoichiometric binary oxide which displays superconducting behavior, and given the pervasive use of Nb in various superconducting circuits, understanding the physics of this compound in thin films is timely and technologically relevant~\cite{mcrae2020materials,siddiqi2021engineering}.

\section*{Methods}
\textit{Molecular beam epitaxy}: The NbO thin films were grown using a Veeco GEN 10 MBE system with a CO$_{2}$ laser substrate heater featuring a 10.6 $\mu$m wavelength laser at the NSF PARADIM facility. Prior to growth, the MgO (001), MgAl$_{2}$O$_{4}$ (001), MgAl$_{2}$O$_{4}$ (111), and Al$_{2}$O$_{3}$ (0001) substrates were laser annealed in situ under ultra-high vacuum conditions at 1150 \dC~ for 200 s, 1550 \dC~ for 300 s, 1600 \dC~ for 200 s, and 1500 \dC~ for 200 s respectively. A Nb flux of approximately $3.2 \cdot 10^{13}$ \textit{atoms}/$cm^{2} \cdot s$ (as measured by a quartz crystal microbalance) was supplied using an electron beam evaporator.  An electron impact emission spectroscopy system was used to monitor the Nb flux during growth. Molecular oxygen was delivered to the substrate through a leak valve and monitored using a baratron. 

\textit{Electron microscopy}: Cross-sectional specimens for scanning transmission electron microscopy (STEM) were prepared using a focused ion beam (FIB) system (FEI Helios NanoLab 600 DualBeam). Samples were initially milled at 30 kV, followed by final thinning and polishing at 5 kV. STEM imaging was conducted using a Thermo Fisher Scientific Themis-Z microscope operated at 200 kV with a probe semi-convergence angle of 20 mrad. High-angle annular dark-field (HAADF) images were collected using a detector range of 64–200 mrad. High-resolution images were acquired by averaging 20 rapid scans (2048 × 2048 pixels, 250 ns dwell time per pixel) to enhance the signal-to-noise ratio. Energy-dispersive X-ray spectroscopy (EDS) was performed using a Super-X detector, and elemental distributions are presented as net count maps. 

\textit{X-ray diffraction}: X-ray diffraction analysis was performed using a Rigaku Smartlab diffractometer with a Cu-$K\alpha1$ source. Reciprocal space mapping data was obtained using the Rigaku HyPix 3000 1D detector.

\textit{Electrical transport}: Electrical properties were measured using a Quantum Design Dynacool system with a dilution refrigerator option. Aluminum wire contacts were made in a Van der Pauw configuration for resistivity measurements of the sample. Thickness values were obtained by fitting the X-ray diffraction data.

\section*{Acknowledgments}
We thank Lawrence Qiu, Tomas Kraay, Brendan D. Faeth, Darrell G. Schlom and Bharat Jalan for their technical support and valuable comments. We acknowledge funding provided by the Gordon and Betty Moore Foundation’s EPiQS Initiative (Grant number GBMF10638 and ICAM-I2CAM, Grant number GBMF5305), and the Institute for Quantum Information and Matter, a NSF Physics Frontiers Center (NSF Grant PHY-2317110). The materials in this paper were grown and characterized using the PARADIM (https://www.paradim.org) Thin Film Facility at Cornell University. PARADIM is supported by the National Science Foundation as a Materials Innovation Platform under Cooperative Agreement DMR-1539918. We acknowledge the Beckman Institute for their support of the X-Ray Crystallography Facility at Caltech. Electron microscopy was performed at the Center for Electron Microscopy and Analysis (CEMAS) at The Ohio State University. S.S.-R. acknowledges Dan Huber for assistance with TEM specimen preparation.
\clearpage

\section*{}

\begin{thebibliography}{42}%
\makeatletter
\providecommand \@ifxundefined [1]{%
 \@ifx{#1\undefined}
}%
\providecommand \@ifnum [1]{%
 \ifnum #1\expandafter \@firstoftwo
 \else \expandafter \@secondoftwo
 \fi
}%
\providecommand \@ifx [1]{%
 \ifx #1\expandafter \@firstoftwo
 \else \expandafter \@secondoftwo
 \fi
}%
\providecommand \natexlab [1]{#1}%
\providecommand \enquote  [1]{``#1''}%
\providecommand \bibnamefont  [1]{#1}%
\providecommand \bibfnamefont [1]{#1}%
\providecommand \citenamefont [1]{#1}%
\providecommand \href@noop [0]{\@secondoftwo}%
\providecommand \href [0]{\begingroup \@sanitize@url \@href}%
\providecommand \@href[1]{\@@startlink{#1}\@@href}%
\providecommand \@@href[1]{\endgroup#1\@@endlink}%
\providecommand \@sanitize@url [0]{\catcode `\\12\catcode `\$12\catcode `\&12\catcode `\#12\catcode `\^12\catcode `\_12\catcode `\%12\relax}%
\providecommand \@@startlink[1]{}%
\providecommand \@@endlink[0]{}%
\providecommand \url  [0]{\begingroup\@sanitize@url \@url }%
\providecommand \@url [1]{\endgroup\@href {#1}{\urlprefix }}%
\providecommand \urlprefix  [0]{URL }%
\providecommand \Eprint [0]{\href }%
\providecommand \doibase [0]{https://doi.org/}%
\providecommand \selectlanguage [0]{\@gobble}%
\providecommand \bibinfo  [0]{\@secondoftwo}%
\providecommand \bibfield  [0]{\@secondoftwo}%
\providecommand \translation [1]{[#1]}%
\providecommand \BibitemOpen [0]{}%
\providecommand \bibitemStop [0]{}%
\providecommand \bibitemNoStop [0]{.\EOS\space}%
\providecommand \EOS [0]{\spacefactor3000\relax}%
\providecommand \BibitemShut  [1]{\csname bibitem#1\endcsname}%
\let\auto@bib@innerbib\@empty
\bibitem [{\citenamefont {Khomskii}(2014)}]{khomskii2014transition}%
  \BibitemOpen
  \bibfield  {author} {\bibinfo {author} {\bibfnamefont {D.~I.}\ \bibnamefont {Khomskii}},\ }\href {https://doi.org/10.1017/CBO9781139096782} {\emph {\bibinfo {title} {Transition metal compounds}}}\ (\bibinfo  {publisher} {Cambridge University Press},\ \bibinfo {year} {2014})\BibitemShut {NoStop}%
\bibitem [{\citenamefont {Imada}\ \emph {et~al.}(1998)\citenamefont {Imada}, \citenamefont {Fujimori},\ and\ \citenamefont {Tokura}}]{imadaMetalinsulatorTransitions1998b}%
  \BibitemOpen
  \bibfield  {author} {\bibinfo {author} {\bibfnamefont {M.}~\bibnamefont {Imada}}, \bibinfo {author} {\bibfnamefont {A.}~\bibnamefont {Fujimori}},\ and\ \bibinfo {author} {\bibfnamefont {Y.}~\bibnamefont {Tokura}},\ }\bibfield  {title} {\bibinfo {title} {Metal-insulator transitions},\ }\href {https://doi.org/10.1103/RevModPhys.70.1039} {\bibfield  {journal} {\bibinfo  {journal} {Reviews of Modern Physics}\ }\textbf {\bibinfo {volume} {70}},\ \bibinfo {pages} {1039} (\bibinfo {year} {1998})}\BibitemShut {NoStop}%
\bibitem [{\citenamefont {Zaanen}\ \emph {et~al.}(1985)\citenamefont {Zaanen}, \citenamefont {Sawatzky},\ and\ \citenamefont {Allen}}]{zaanen1985band}%
  \BibitemOpen
  \bibfield  {author} {\bibinfo {author} {\bibfnamefont {J.}~\bibnamefont {Zaanen}}, \bibinfo {author} {\bibfnamefont {G.}~\bibnamefont {Sawatzky}},\ and\ \bibinfo {author} {\bibfnamefont {J.}~\bibnamefont {Allen}},\ }\bibfield  {title} {\bibinfo {title} {Band gaps and electronic structure of transition-metal compounds},\ }\href {https://doi.org/10.1103/PhysRevLett.55.418} {\bibfield  {journal} {\bibinfo  {journal} {Physical review letters}\ }\textbf {\bibinfo {volume} {55}},\ \bibinfo {pages} {418} (\bibinfo {year} {1985})}\BibitemShut {NoStop}%
\bibitem [{\citenamefont {Cox}\ and\ \citenamefont {Sleight}(1976)}]{cox1976crystal}%
  \BibitemOpen
  \bibfield  {author} {\bibinfo {author} {\bibfnamefont {D.}~\bibnamefont {Cox}}\ and\ \bibinfo {author} {\bibfnamefont {A.}~\bibnamefont {Sleight}},\ }\bibfield  {title} {\bibinfo {title} {Crystal structure of ba2bi3+ bi5+ o6},\ }\href {https://doi.org/10.1016/0038-1098(76)90632-3} {\bibfield  {journal} {\bibinfo  {journal} {Solid State Communications}\ }\textbf {\bibinfo {volume} {19}},\ \bibinfo {pages} {969} (\bibinfo {year} {1976})}\BibitemShut {NoStop}%
\bibitem [{\citenamefont {Johnston}\ \emph {et~al.}(2014)\citenamefont {Johnston}, \citenamefont {Mukherjee}, \citenamefont {Elfimov}, \citenamefont {Berciu},\ and\ \citenamefont {Sawatzky}}]{johnston2014charge}%
  \BibitemOpen
  \bibfield  {author} {\bibinfo {author} {\bibfnamefont {S.}~\bibnamefont {Johnston}}, \bibinfo {author} {\bibfnamefont {A.}~\bibnamefont {Mukherjee}}, \bibinfo {author} {\bibfnamefont {I.}~\bibnamefont {Elfimov}}, \bibinfo {author} {\bibfnamefont {M.}~\bibnamefont {Berciu}},\ and\ \bibinfo {author} {\bibfnamefont {G.~A.}\ \bibnamefont {Sawatzky}},\ }\bibfield  {title} {\bibinfo {title} {Charge disproportionation without charge transfer in the rare-earth-element nickelates as a possible mechanism for the metal-insulator transition},\ }\href {https://doi.org/10.1103/PhysRevLett.112.106404} {\bibfield  {journal} {\bibinfo  {journal} {Physical review letters}\ }\textbf {\bibinfo {volume} {112}},\ \bibinfo {pages} {106404} (\bibinfo {year} {2014})}\BibitemShut {NoStop}%
\bibitem [{\citenamefont {Sun}\ \emph {et~al.}(2023)\citenamefont {Sun}, \citenamefont {Huo}, \citenamefont {Hu}, \citenamefont {Li}, \citenamefont {Liu}, \citenamefont {Han}, \citenamefont {Tang}, \citenamefont {Mao}, \citenamefont {Yang}, \citenamefont {Wang} \emph {et~al.}}]{sun2023signatures}%
  \BibitemOpen
  \bibfield  {author} {\bibinfo {author} {\bibfnamefont {H.}~\bibnamefont {Sun}}, \bibinfo {author} {\bibfnamefont {M.}~\bibnamefont {Huo}}, \bibinfo {author} {\bibfnamefont {X.}~\bibnamefont {Hu}}, \bibinfo {author} {\bibfnamefont {J.}~\bibnamefont {Li}}, \bibinfo {author} {\bibfnamefont {Z.}~\bibnamefont {Liu}}, \bibinfo {author} {\bibfnamefont {Y.}~\bibnamefont {Han}}, \bibinfo {author} {\bibfnamefont {L.}~\bibnamefont {Tang}}, \bibinfo {author} {\bibfnamefont {Z.}~\bibnamefont {Mao}}, \bibinfo {author} {\bibfnamefont {P.}~\bibnamefont {Yang}}, \bibinfo {author} {\bibfnamefont {B.}~\bibnamefont {Wang}}, \emph {et~al.},\ }\bibfield  {title} {\bibinfo {title} {Signatures of superconductivity near 80 k in a nickelate under high pressure},\ }\href {https://doi.org/10.1038/s41586-023-06408-7} {\bibfield  {journal} {\bibinfo  {journal} {Nature}\ }\textbf {\bibinfo {volume} {621}},\ \bibinfo {pages} {493} (\bibinfo {year} {2023})}\BibitemShut {NoStop}%
\bibitem [{\citenamefont {Wang}\ \emph {et~al.}(2020)\citenamefont {Wang}, \citenamefont {Zhang}, \citenamefont {Zhu}, \citenamefont {L{\"u}}, \citenamefont {Li}, \citenamefont {Zou},\ and\ \citenamefont {Zhao}}]{wang2020antiperovskites}%
  \BibitemOpen
  \bibfield  {author} {\bibinfo {author} {\bibfnamefont {Y.}~\bibnamefont {Wang}}, \bibinfo {author} {\bibfnamefont {H.}~\bibnamefont {Zhang}}, \bibinfo {author} {\bibfnamefont {J.}~\bibnamefont {Zhu}}, \bibinfo {author} {\bibfnamefont {X.}~\bibnamefont {L{\"u}}}, \bibinfo {author} {\bibfnamefont {S.}~\bibnamefont {Li}}, \bibinfo {author} {\bibfnamefont {R.}~\bibnamefont {Zou}},\ and\ \bibinfo {author} {\bibfnamefont {Y.}~\bibnamefont {Zhao}},\ }\bibfield  {title} {\bibinfo {title} {Antiperovskites with exceptional functionalities},\ }\href {https://doi.org/10.1002/adma.201905007} {\bibfield  {journal} {\bibinfo  {journal} {Advanced Materials}\ }\textbf {\bibinfo {volume} {32}},\ \bibinfo {pages} {1905007} (\bibinfo {year} {2020})}\BibitemShut {NoStop}%
\bibitem [{\citenamefont {Li}\ \emph {et~al.}(2019)\citenamefont {Li}, \citenamefont {Lee}, \citenamefont {Wang}, \citenamefont {Osada}, \citenamefont {Crossley}, \citenamefont {Lee}, \citenamefont {Cui}, \citenamefont {Hikita},\ and\ \citenamefont {Hwang}}]{li2019superconductivity}%
  \BibitemOpen
  \bibfield  {author} {\bibinfo {author} {\bibfnamefont {D.}~\bibnamefont {Li}}, \bibinfo {author} {\bibfnamefont {K.}~\bibnamefont {Lee}}, \bibinfo {author} {\bibfnamefont {B.~Y.}\ \bibnamefont {Wang}}, \bibinfo {author} {\bibfnamefont {M.}~\bibnamefont {Osada}}, \bibinfo {author} {\bibfnamefont {S.}~\bibnamefont {Crossley}}, \bibinfo {author} {\bibfnamefont {H.~R.}\ \bibnamefont {Lee}}, \bibinfo {author} {\bibfnamefont {Y.}~\bibnamefont {Cui}}, \bibinfo {author} {\bibfnamefont {Y.}~\bibnamefont {Hikita}},\ and\ \bibinfo {author} {\bibfnamefont {H.~Y.}\ \bibnamefont {Hwang}},\ }\bibfield  {title} {\bibinfo {title} {Superconductivity in an infinite-layer nickelate},\ }\href {https://doi.org/10.1038/s41586-019-1496-5} {\bibfield  {journal} {\bibinfo  {journal} {Nature}\ }\textbf {\bibinfo {volume} {572}},\ \bibinfo {pages} {624} (\bibinfo {year} {2019})}\BibitemShut {NoStop}%
\bibitem [{\citenamefont {Kim}\ \emph {et~al.}(2023)\citenamefont {Kim}, \citenamefont {Smeaton}, \citenamefont {Jia}, \citenamefont {Goodge}, \citenamefont {Cho}, \citenamefont {Lee}, \citenamefont {Osada}, \citenamefont {Jost}, \citenamefont {Ievlev}, \citenamefont {Moritz} \emph {et~al.}}]{kim2023geometric}%
  \BibitemOpen
  \bibfield  {author} {\bibinfo {author} {\bibfnamefont {W.~J.}\ \bibnamefont {Kim}}, \bibinfo {author} {\bibfnamefont {M.~A.}\ \bibnamefont {Smeaton}}, \bibinfo {author} {\bibfnamefont {C.}~\bibnamefont {Jia}}, \bibinfo {author} {\bibfnamefont {B.~H.}\ \bibnamefont {Goodge}}, \bibinfo {author} {\bibfnamefont {B.-G.}\ \bibnamefont {Cho}}, \bibinfo {author} {\bibfnamefont {K.}~\bibnamefont {Lee}}, \bibinfo {author} {\bibfnamefont {M.}~\bibnamefont {Osada}}, \bibinfo {author} {\bibfnamefont {D.}~\bibnamefont {Jost}}, \bibinfo {author} {\bibfnamefont {A.~V.}\ \bibnamefont {Ievlev}}, \bibinfo {author} {\bibfnamefont {B.}~\bibnamefont {Moritz}}, \emph {et~al.},\ }\bibfield  {title} {\bibinfo {title} {Geometric frustration of jahn--teller order in the infinite-layer lattice},\ }\href {https://doi.org/10.1038/s41586-023-06432-7} {\bibfield  {journal} {\bibinfo  {journal} {Nature}\ }\textbf {\bibinfo {volume} {615}},\ \bibinfo {pages} {237} (\bibinfo {year} {2023})}\BibitemShut {NoStop}%
\bibitem [{\citenamefont {Ko}\ \emph {et~al.}(2024)\citenamefont {Ko}, \citenamefont {Yu}, \citenamefont {Liu}, \citenamefont {Bhatt}, \citenamefont {Li}, \citenamefont {Thampy}, \citenamefont {Kuo}, \citenamefont {Wang}, \citenamefont {Lee}, \citenamefont {Lee} \emph {et~al.}}]{ko2024signatures}%
  \BibitemOpen
  \bibfield  {author} {\bibinfo {author} {\bibfnamefont {E.~K.}\ \bibnamefont {Ko}}, \bibinfo {author} {\bibfnamefont {Y.}~\bibnamefont {Yu}}, \bibinfo {author} {\bibfnamefont {Y.}~\bibnamefont {Liu}}, \bibinfo {author} {\bibfnamefont {L.}~\bibnamefont {Bhatt}}, \bibinfo {author} {\bibfnamefont {J.}~\bibnamefont {Li}}, \bibinfo {author} {\bibfnamefont {V.}~\bibnamefont {Thampy}}, \bibinfo {author} {\bibfnamefont {C.-T.}\ \bibnamefont {Kuo}}, \bibinfo {author} {\bibfnamefont {B.~Y.}\ \bibnamefont {Wang}}, \bibinfo {author} {\bibfnamefont {Y.}~\bibnamefont {Lee}}, \bibinfo {author} {\bibfnamefont {K.}~\bibnamefont {Lee}}, \emph {et~al.},\ }\bibfield  {title} {\bibinfo {title} {Signatures of ambient pressure superconductivity in thin film la3ni2o7},\ }\href {https://doi.org/10.1038/s41586-024-08525-3} {\bibfield  {journal} {\bibinfo  {journal} {Nature}\ ,\ \bibinfo {pages} {1}} (\bibinfo {year} {2024})}\BibitemShut {NoStop}%
\bibitem [{\citenamefont {Mairoser}\ \emph {et~al.}(2015)\citenamefont {Mairoser}, \citenamefont {Mundy}, \citenamefont {Melville}, \citenamefont {Hodash}, \citenamefont {Cueva}, \citenamefont {Held}, \citenamefont {Glavic}, \citenamefont {Schubert}, \citenamefont {Muller}, \citenamefont {Schlom} \emph {et~al.}}]{mairoser2015high}%
  \BibitemOpen
  \bibfield  {author} {\bibinfo {author} {\bibfnamefont {T.}~\bibnamefont {Mairoser}}, \bibinfo {author} {\bibfnamefont {J.~A.}\ \bibnamefont {Mundy}}, \bibinfo {author} {\bibfnamefont {A.}~\bibnamefont {Melville}}, \bibinfo {author} {\bibfnamefont {D.}~\bibnamefont {Hodash}}, \bibinfo {author} {\bibfnamefont {P.}~\bibnamefont {Cueva}}, \bibinfo {author} {\bibfnamefont {R.}~\bibnamefont {Held}}, \bibinfo {author} {\bibfnamefont {A.}~\bibnamefont {Glavic}}, \bibinfo {author} {\bibfnamefont {J.}~\bibnamefont {Schubert}}, \bibinfo {author} {\bibfnamefont {D.~A.}\ \bibnamefont {Muller}}, \bibinfo {author} {\bibfnamefont {D.~G.}\ \bibnamefont {Schlom}}, \emph {et~al.},\ }\bibfield  {title} {\bibinfo {title} {High-quality euo thin films the easy way via topotactic transformation},\ }\href {https://doi.org/10.1038/ncomms8716} {\bibfield  {journal} {\bibinfo  {journal} {Nature communications}\ }\textbf {\bibinfo {volume} {6}},\ \bibinfo {pages} {7716} (\bibinfo {year} {2015})}\BibitemShut {NoStop}%
\bibitem [{\citenamefont {Jeen}\ \emph {et~al.}(2013)\citenamefont {Jeen}, \citenamefont {Choi}, \citenamefont {Biegalski}, \citenamefont {Folkman}, \citenamefont {Tung}, \citenamefont {Fong}, \citenamefont {Freeland}, \citenamefont {Shin}, \citenamefont {Ohta}, \citenamefont {Chisholm} \emph {et~al.}}]{jeen2013reversible}%
  \BibitemOpen
  \bibfield  {author} {\bibinfo {author} {\bibfnamefont {H.}~\bibnamefont {Jeen}}, \bibinfo {author} {\bibfnamefont {W.~S.}\ \bibnamefont {Choi}}, \bibinfo {author} {\bibfnamefont {M.~D.}\ \bibnamefont {Biegalski}}, \bibinfo {author} {\bibfnamefont {C.~M.}\ \bibnamefont {Folkman}}, \bibinfo {author} {\bibfnamefont {I.-C.}\ \bibnamefont {Tung}}, \bibinfo {author} {\bibfnamefont {D.~D.}\ \bibnamefont {Fong}}, \bibinfo {author} {\bibfnamefont {J.~W.}\ \bibnamefont {Freeland}}, \bibinfo {author} {\bibfnamefont {D.}~\bibnamefont {Shin}}, \bibinfo {author} {\bibfnamefont {H.}~\bibnamefont {Ohta}}, \bibinfo {author} {\bibfnamefont {M.~F.}\ \bibnamefont {Chisholm}}, \emph {et~al.},\ }\bibfield  {title} {\bibinfo {title} {Reversible redox reactions in an epitaxially stabilized srcoo x oxygen sponge},\ }\href {https://doi.org/10.1038/nmat3736} {\bibfield  {journal} {\bibinfo  {journal} {Nature materials}\ }\textbf {\bibinfo {volume} {12}},\ \bibinfo {pages} {1057} (\bibinfo {year} {2013})}\BibitemShut {NoStop}%
\bibitem [{\citenamefont {Hulm}\ \emph {et~al.}(1972)\citenamefont {Hulm}, \citenamefont {Jones}, \citenamefont {Hein},\ and\ \citenamefont {Gibson}}]{hulm1972superconductivity}%
  \BibitemOpen
  \bibfield  {author} {\bibinfo {author} {\bibfnamefont {J.}~\bibnamefont {Hulm}}, \bibinfo {author} {\bibfnamefont {C.}~\bibnamefont {Jones}}, \bibinfo {author} {\bibfnamefont {R.}~\bibnamefont {Hein}},\ and\ \bibinfo {author} {\bibfnamefont {J.}~\bibnamefont {Gibson}},\ }\bibfield  {title} {\bibinfo {title} {Superconductivity in the tio and nbo systems},\ }\href {https://doi.org/10.1007/BF00660068} {\bibfield  {journal} {\bibinfo  {journal} {Journal of Low Temperature Physics}\ }\textbf {\bibinfo {volume} {7}},\ \bibinfo {pages} {291} (\bibinfo {year} {1972})}\BibitemShut {NoStop}%
\bibitem [{\citenamefont {Okaz}\ and\ \citenamefont {Keesom}(1975)}]{okaz1975specific}%
  \BibitemOpen
  \bibfield  {author} {\bibinfo {author} {\bibfnamefont {A.~M.}\ \bibnamefont {Okaz}}\ and\ \bibinfo {author} {\bibfnamefont {P.}~\bibnamefont {Keesom}},\ }\bibfield  {title} {\bibinfo {title} {Specific heat and magnetization of the superconducting monoxides: Nbo and tio},\ }\href {https://doi.org/10.1103/PhysRevB.12.4917} {\bibfield  {journal} {\bibinfo  {journal} {Physical Review B}\ }\textbf {\bibinfo {volume} {12}},\ \bibinfo {pages} {4917} (\bibinfo {year} {1975})}\BibitemShut {NoStop}%
\bibitem [{\citenamefont {Efimenko}\ \emph {et~al.}(2017)\citenamefont {Efimenko}, \citenamefont {Hollmann}, \citenamefont {Hoefer}, \citenamefont {Weinen}, \citenamefont {Takegami}, \citenamefont {Wolff}, \citenamefont {Altendorf}, \citenamefont {Hu}, \citenamefont {Rata}, \citenamefont {Komarek} \emph {et~al.}}]{efimenko2017electronic}%
  \BibitemOpen
  \bibfield  {author} {\bibinfo {author} {\bibfnamefont {A.~K.}\ \bibnamefont {Efimenko}}, \bibinfo {author} {\bibfnamefont {N.}~\bibnamefont {Hollmann}}, \bibinfo {author} {\bibfnamefont {K.}~\bibnamefont {Hoefer}}, \bibinfo {author} {\bibfnamefont {J.}~\bibnamefont {Weinen}}, \bibinfo {author} {\bibfnamefont {D.}~\bibnamefont {Takegami}}, \bibinfo {author} {\bibfnamefont {K.~K.}\ \bibnamefont {Wolff}}, \bibinfo {author} {\bibfnamefont {S.~G.}\ \bibnamefont {Altendorf}}, \bibinfo {author} {\bibfnamefont {Z.}~\bibnamefont {Hu}}, \bibinfo {author} {\bibfnamefont {A.~D.}\ \bibnamefont {Rata}}, \bibinfo {author} {\bibfnamefont {A.~C.}\ \bibnamefont {Komarek}}, \emph {et~al.},\ }\bibfield  {title} {\bibinfo {title} {Electronic signature of the vacancy ordering in nbo (nb 3 o 3)},\ }\href {https://doi.org/10.1103/PhysRevB.96.195112} {\bibfield  {journal} {\bibinfo  {journal} {Physical Review B}\ }\textbf {\bibinfo {volume} {96}},\ \bibinfo {pages} {195112} (\bibinfo {year} {2017})}\BibitemShut {NoStop}%
\bibitem [{\citenamefont {Burdett}\ and\ \citenamefont {Hughbanks}(1984)}]{burdett1984niobium}%
  \BibitemOpen
  \bibfield  {author} {\bibinfo {author} {\bibfnamefont {J.~K.}\ \bibnamefont {Burdett}}\ and\ \bibinfo {author} {\bibfnamefont {T.}~\bibnamefont {Hughbanks}},\ }\bibfield  {title} {\bibinfo {title} {Niobium oxide (nbo) and titanium oxide (tio): a study of the structural and electronic stability of structures derived from rock salt},\ }\href {https://doi.org/10.1021/ja00323a006} {\bibfield  {journal} {\bibinfo  {journal} {Journal of the American Chemical Society}\ }\textbf {\bibinfo {volume} {106}},\ \bibinfo {pages} {3101} (\bibinfo {year} {1984})}\BibitemShut {NoStop}%
\bibitem [{\citenamefont {Liu}\ \emph {et~al.}(2014)\citenamefont {Liu}, \citenamefont {Zhou}, \citenamefont {Gall},\ and\ \citenamefont {Khare}}]{liu2014first}%
  \BibitemOpen
  \bibfield  {author} {\bibinfo {author} {\bibfnamefont {Z.}~\bibnamefont {Liu}}, \bibinfo {author} {\bibfnamefont {X.}~\bibnamefont {Zhou}}, \bibinfo {author} {\bibfnamefont {D.}~\bibnamefont {Gall}},\ and\ \bibinfo {author} {\bibfnamefont {S.}~\bibnamefont {Khare}},\ }\bibfield  {title} {\bibinfo {title} {First-principles investigation of the structural, mechanical and electronic properties of the nbo-structured 3d, 4d and 5d transition metal nitrides},\ }\href {https://doi.org/10.1016/j.commatsci.2013.12.038} {\bibfield  {journal} {\bibinfo  {journal} {Computational materials science}\ }\textbf {\bibinfo {volume} {84}},\ \bibinfo {pages} {365} (\bibinfo {year} {2014})}\BibitemShut {NoStop}%
\bibitem [{\citenamefont {Zhou}\ \emph {et~al.}(2022)\citenamefont {Zhou}, \citenamefont {Gu}, \citenamefont {Song}, \citenamefont {Ma}, \citenamefont {Fang}, \citenamefont {Zhou}, \citenamefont {Zhao}, \citenamefont {Liang},\ and\ \citenamefont {Wang}}]{zhou2022synthesis}%
  \BibitemOpen
  \bibfield  {author} {\bibinfo {author} {\bibfnamefont {X.}~\bibnamefont {Zhou}}, \bibinfo {author} {\bibfnamefont {C.}~\bibnamefont {Gu}}, \bibinfo {author} {\bibfnamefont {G.}~\bibnamefont {Song}}, \bibinfo {author} {\bibfnamefont {D.}~\bibnamefont {Ma}}, \bibinfo {author} {\bibfnamefont {L.}~\bibnamefont {Fang}}, \bibinfo {author} {\bibfnamefont {C.}~\bibnamefont {Zhou}}, \bibinfo {author} {\bibfnamefont {Y.}~\bibnamefont {Zhao}}, \bibinfo {author} {\bibfnamefont {Y.}~\bibnamefont {Liang}},\ and\ \bibinfo {author} {\bibfnamefont {S.}~\bibnamefont {Wang}},\ }\bibfield  {title} {\bibinfo {title} {Synthesis, crystal structures, mechanical properties, and formation mechanisms of cubic tungsten nitrides},\ }\href {https://doi.org/10.1021/acs.chemmater.2c02563} {\bibfield  {journal} {\bibinfo  {journal} {Chemistry of Materials}\ }\textbf {\bibinfo {volume} {34}},\ \bibinfo {pages} {9261} (\bibinfo {year} {2022})}\BibitemShut {NoStop}%
\bibitem [{\citenamefont {Janninck}\ and\ \citenamefont {Whitmore}(1966)}]{janninck1966electrical}%
  \BibitemOpen
  \bibfield  {author} {\bibinfo {author} {\bibfnamefont {R.}~\bibnamefont {Janninck}}\ and\ \bibinfo {author} {\bibfnamefont {D.}~\bibnamefont {Whitmore}},\ }\bibfield  {title} {\bibinfo {title} {Electrical conductivity and thermoelectric power of niobium dioxide},\ }\href {https://doi.org/10.1016/0022-3697(66)90094-1} {\bibfield  {journal} {\bibinfo  {journal} {Journal of Physics and Chemistry of Solids}\ }\textbf {\bibinfo {volume} {27}},\ \bibinfo {pages} {1183} (\bibinfo {year} {1966})}\BibitemShut {NoStop}%
\bibitem [{\citenamefont {Valeeva}\ \emph {et~al.}(2023)\citenamefont {Valeeva}, \citenamefont {Rempel}, \citenamefont {Naumov}, \citenamefont {Kuranova},\ and\ \citenamefont {Sсhrottner}}]{valeeva2023niobium}%
  \BibitemOpen
  \bibfield  {author} {\bibinfo {author} {\bibfnamefont {Ð.}~\bibnamefont {Valeeva}}, \bibinfo {author} {\bibfnamefont {Ð.}~\bibnamefont {Rempel}}, \bibinfo {author} {\bibfnamefont {S.}~\bibnamefont {Naumov}}, \bibinfo {author} {\bibfnamefont {N.}~\bibnamefont {Kuranova}},\ and\ \bibinfo {author} {\bibfnamefont {H.}~\bibnamefont {Sсhrottner}},\ }\bibfield  {title} {\bibinfo {title} {Niobium monoxide nbo single crystal growing by floating zone method},\ }\href {https://doi.org/10.1016/j.matchar.2023.113265} {\bibfield  {journal} {\bibinfo  {journal} {Materials Characterization}\ }\textbf {\bibinfo {volume} {205}},\ \bibinfo {pages} {113265} (\bibinfo {year} {2023})}\BibitemShut {NoStop}%
\bibitem [{\citenamefont {Kim}\ \emph {et~al.}(2025)\citenamefont {Kim}, \citenamefont {Glotzer}, \citenamefont {Llanos}, \citenamefont {Salmani-Rezaie},\ and\ \citenamefont {Falson}}]{kim2025high}%
  \BibitemOpen
  \bibfield  {author} {\bibinfo {author} {\bibfnamefont {J.~R.}\ \bibnamefont {Kim}}, \bibinfo {author} {\bibfnamefont {S.}~\bibnamefont {Glotzer}}, \bibinfo {author} {\bibfnamefont {A.}~\bibnamefont {Llanos}}, \bibinfo {author} {\bibfnamefont {S.}~\bibnamefont {Salmani-Rezaie}},\ and\ \bibinfo {author} {\bibfnamefont {J.}~\bibnamefont {Falson}},\ }\bibfield  {title} {\bibinfo {title} {High-temperature diffusion enabled epitaxy of the ti--o system},\ }\href {https://doi.org/10.1002/adma.202413447} {\bibfield  {journal} {\bibinfo  {journal} {Advanced Materials}\ }\textbf {\bibinfo {volume} {37}},\ \bibinfo {pages} {2413447} (\bibinfo {year} {2025})}\BibitemShut {NoStop}%
\bibitem [{\citenamefont {Gao}\ \emph {et~al.}(1998)\citenamefont {Gao}, \citenamefont {Kim},\ and\ \citenamefont {Chambers}}]{gao1998preparation}%
  \BibitemOpen
  \bibfield  {author} {\bibinfo {author} {\bibfnamefont {Y.}~\bibnamefont {Gao}}, \bibinfo {author} {\bibfnamefont {Y.~J.}\ \bibnamefont {Kim}},\ and\ \bibinfo {author} {\bibfnamefont {S.~A.}\ \bibnamefont {Chambers}},\ }\bibfield  {title} {\bibinfo {title} {Preparation and characterization of epitaxial iron oxide films},\ }\href {https://doi.org/10.1557/JMR.1998.0281} {\bibfield  {journal} {\bibinfo  {journal} {Journal of materials research}\ }\textbf {\bibinfo {volume} {13}},\ \bibinfo {pages} {2003} (\bibinfo {year} {1998})}\BibitemShut {NoStop}%
\bibitem [{\citenamefont {Rata}\ \emph {et~al.}(2004)\citenamefont {Rata}, \citenamefont {Chezan}, \citenamefont {Haverkort}, \citenamefont {Hsieh}, \citenamefont {Lin}, \citenamefont {Chen}, \citenamefont {Tjeng},\ and\ \citenamefont {Hibma}}]{rata2004growth}%
  \BibitemOpen
  \bibfield  {author} {\bibinfo {author} {\bibfnamefont {A.}~\bibnamefont {Rata}}, \bibinfo {author} {\bibfnamefont {A.}~\bibnamefont {Chezan}}, \bibinfo {author} {\bibfnamefont {M.}~\bibnamefont {Haverkort}}, \bibinfo {author} {\bibfnamefont {H.}~\bibnamefont {Hsieh}}, \bibinfo {author} {\bibfnamefont {H.-J.}\ \bibnamefont {Lin}}, \bibinfo {author} {\bibfnamefont {C.}~\bibnamefont {Chen}}, \bibinfo {author} {\bibfnamefont {L.}~\bibnamefont {Tjeng}},\ and\ \bibinfo {author} {\bibfnamefont {T.}~\bibnamefont {Hibma}},\ }\bibfield  {title} {\bibinfo {title} {Growth and properties of strained vo x thin films with controlled stoichiometry},\ }\href {https://doi.org/10.1103/PhysRevB.69.075404} {\bibfield  {journal} {\bibinfo  {journal} {Physical Review B}\ }\textbf {\bibinfo {volume} {69}},\ \bibinfo {pages} {075404} (\bibinfo {year} {2004})}\BibitemShut {NoStop}%
\bibitem [{\citenamefont {Budde}\ \emph {et~al.}(2018)\citenamefont {Budde}, \citenamefont {Tschammer}, \citenamefont {Franz}, \citenamefont {Feldl}, \citenamefont {Ramsteiner}, \citenamefont {Goldhahn}, \citenamefont {Feneberg}, \citenamefont {Barsan}, \citenamefont {Oprea},\ and\ \citenamefont {Bierwagen}}]{budde2018structural}%
  \BibitemOpen
  \bibfield  {author} {\bibinfo {author} {\bibfnamefont {M.}~\bibnamefont {Budde}}, \bibinfo {author} {\bibfnamefont {C.}~\bibnamefont {Tschammer}}, \bibinfo {author} {\bibfnamefont {P.}~\bibnamefont {Franz}}, \bibinfo {author} {\bibfnamefont {J.}~\bibnamefont {Feldl}}, \bibinfo {author} {\bibfnamefont {M.}~\bibnamefont {Ramsteiner}}, \bibinfo {author} {\bibfnamefont {R.}~\bibnamefont {Goldhahn}}, \bibinfo {author} {\bibfnamefont {M.}~\bibnamefont {Feneberg}}, \bibinfo {author} {\bibfnamefont {N.}~\bibnamefont {Barsan}}, \bibinfo {author} {\bibfnamefont {A.}~\bibnamefont {Oprea}},\ and\ \bibinfo {author} {\bibfnamefont {O.}~\bibnamefont {Bierwagen}},\ }\bibfield  {title} {\bibinfo {title} {Structural, optical, and electrical properties of unintentionally doped nio layers grown on mgo by plasma-assisted molecular beam epitaxy},\ }\bibfield  {journal} {\bibinfo  {journal} {Journal of Applied Physics}\ }\textbf {\bibinfo {volume} {123}},\ \href {https://doi.org/10.1063/1.5026738} {10.1063/1.5026738} (\bibinfo
  {year} {2018})\BibitemShut {NoStop}%
\bibitem [{\citenamefont {Li}\ \emph {et~al.}(2021)\citenamefont {Li}, \citenamefont {Zou}, \citenamefont {Han}, \citenamefont {Foyevtsova}, \citenamefont {Shin}, \citenamefont {Lee}, \citenamefont {Liu}, \citenamefont {Shin}, \citenamefont {Albright}, \citenamefont {Sutarto} \emph {et~al.}}]{li2021single}%
  \BibitemOpen
  \bibfield  {author} {\bibinfo {author} {\bibfnamefont {F.}~\bibnamefont {Li}}, \bibinfo {author} {\bibfnamefont {Y.}~\bibnamefont {Zou}}, \bibinfo {author} {\bibfnamefont {M.-G.}\ \bibnamefont {Han}}, \bibinfo {author} {\bibfnamefont {K.}~\bibnamefont {Foyevtsova}}, \bibinfo {author} {\bibfnamefont {H.}~\bibnamefont {Shin}}, \bibinfo {author} {\bibfnamefont {S.}~\bibnamefont {Lee}}, \bibinfo {author} {\bibfnamefont {C.}~\bibnamefont {Liu}}, \bibinfo {author} {\bibfnamefont {K.}~\bibnamefont {Shin}}, \bibinfo {author} {\bibfnamefont {S.~D.}\ \bibnamefont {Albright}}, \bibinfo {author} {\bibfnamefont {R.}~\bibnamefont {Sutarto}}, \emph {et~al.},\ }\bibfield  {title} {\bibinfo {title} {Single-crystalline epitaxial tio film: A metal and superconductor, similar to ti metal},\ }\href {https://doi.org/10.1126/sciadv.abd4248} {\bibfield  {journal} {\bibinfo  {journal} {Science Advances}\ }\textbf {\bibinfo {volume} {7}},\ \bibinfo {pages} {eabd4248} (\bibinfo {year} {2021})}\BibitemShut {NoStop}%
\bibitem [{\citenamefont {Ananta}(2004)}]{ananta2004synthesis}%
  \BibitemOpen
  \bibfield  {author} {\bibinfo {author} {\bibfnamefont {S.}~\bibnamefont {Ananta}},\ }\bibfield  {title} {\bibinfo {title} {Synthesis, formation and characterization of mg4nb2o9 powders},\ }\href {https://doi.org/10.1016/j.matlet.2004.03.013} {\bibfield  {journal} {\bibinfo  {journal} {Materials Letters}\ }\textbf {\bibinfo {volume} {58}},\ \bibinfo {pages} {2530} (\bibinfo {year} {2004})}\BibitemShut {NoStop}%
\bibitem [{\citenamefont {Kimura}\ \emph {et~al.}(2024)\citenamefont {Kimura}, \citenamefont {Kaminaga}, \citenamefont {Kobayashi}, \citenamefont {Cho}, \citenamefont {Maruyama}, \citenamefont {Maeda},\ and\ \citenamefont {Matsumoto}}]{kimura2024elevating}%
  \BibitemOpen
  \bibfield  {author} {\bibinfo {author} {\bibfnamefont {R.}~\bibnamefont {Kimura}}, \bibinfo {author} {\bibfnamefont {K.}~\bibnamefont {Kaminaga}}, \bibinfo {author} {\bibfnamefont {T.}~\bibnamefont {Kobayashi}}, \bibinfo {author} {\bibfnamefont {Y.}~\bibnamefont {Cho}}, \bibinfo {author} {\bibfnamefont {S.}~\bibnamefont {Maruyama}}, \bibinfo {author} {\bibfnamefont {A.}~\bibnamefont {Maeda}},\ and\ \bibinfo {author} {\bibfnamefont {Y.}~\bibnamefont {Matsumoto}},\ }\bibfield  {title} {\bibinfo {title} {Elevating the superconducting temperature in epitaxially stabilized rock-salt nbo},\ }\href {https://doi.org/10.1021/acs.chemmater.4c00097} {\bibfield  {journal} {\bibinfo  {journal} {Chemistry of Materials}\ }\textbf {\bibinfo {volume} {36}},\ \bibinfo {pages} {5028} (\bibinfo {year} {2024})}\BibitemShut {NoStop}%
\bibitem [{\citenamefont {Prostakova}\ \emph {et~al.}(2012)\citenamefont {Prostakova}, \citenamefont {Chen}, \citenamefont {Jak},\ and\ \citenamefont {Decterov}}]{prostakova2012experimental}%
  \BibitemOpen
  \bibfield  {author} {\bibinfo {author} {\bibfnamefont {V.}~\bibnamefont {Prostakova}}, \bibinfo {author} {\bibfnamefont {J.}~\bibnamefont {Chen}}, \bibinfo {author} {\bibfnamefont {E.}~\bibnamefont {Jak}},\ and\ \bibinfo {author} {\bibfnamefont {S.~A.}\ \bibnamefont {Decterov}},\ }\bibfield  {title} {\bibinfo {title} {Experimental study and thermodynamic optimization of the cao--nio, mgo--nio and nio--sio2 systems},\ }\href {https://doi.org/10.1016/j.calphad.2011.12.009} {\bibfield  {journal} {\bibinfo  {journal} {Calphad}\ }\textbf {\bibinfo {volume} {37}},\ \bibinfo {pages} {1} (\bibinfo {year} {2012})}\BibitemShut {NoStop}%
\bibitem [{\citenamefont {Burnus}\ \emph {et~al.}(1987)\citenamefont {Burnus}, \citenamefont {K{\"o}hler},\ and\ \citenamefont {Simon}}]{burnus1987mn3nb6o11}%
  \BibitemOpen
  \bibfield  {author} {\bibinfo {author} {\bibfnamefont {R.}~\bibnamefont {Burnus}}, \bibinfo {author} {\bibfnamefont {J.}~\bibnamefont {K{\"o}hler}},\ and\ \bibinfo {author} {\bibfnamefont {A.}~\bibnamefont {Simon}},\ }\bibfield  {title} {\bibinfo {title} {Mn3nb6o11 und mg3nb6o11-darstellung von einkristallen und strukturverfeinerung/mn3nb6o11 and mg3nb6o11-preparation of single crystals and structure refinement},\ }\href {https://doi.org/10.1515/znb-1987-0504} {\bibfield  {journal} {\bibinfo  {journal} {Zeitschrift f{\"u}r Naturforschung B}\ }\textbf {\bibinfo {volume} {42}},\ \bibinfo {pages} {536} (\bibinfo {year} {1987})}\BibitemShut {NoStop}%
\bibitem [{\citenamefont {Honig}\ \emph {et~al.}(1973)\citenamefont {Honig}, \citenamefont {Wahnsiedler},\ and\ \citenamefont {Eklund}}]{honig1973electrical}%
  \BibitemOpen
  \bibfield  {author} {\bibinfo {author} {\bibfnamefont {J.}~\bibnamefont {Honig}}, \bibinfo {author} {\bibfnamefont {W.}~\bibnamefont {Wahnsiedler}},\ and\ \bibinfo {author} {\bibfnamefont {P.}~\bibnamefont {Eklund}},\ }\bibfield  {title} {\bibinfo {title} {Electrical properties of nbo in high magnetic fields},\ }\href {https://doi.org/10.1016/0022-4596(73)90183-7} {\bibfield  {journal} {\bibinfo  {journal} {Journal of Solid State Chemistry}\ }\textbf {\bibinfo {volume} {6}},\ \bibinfo {pages} {203} (\bibinfo {year} {1973})}\BibitemShut {NoStop}%
\bibitem [{\citenamefont {Werthamer}\ \emph {et~al.}(1966)\citenamefont {Werthamer}, \citenamefont {Helfand},\ and\ \citenamefont {Hohenberg}}]{werthamer1966temperature}%
  \BibitemOpen
  \bibfield  {author} {\bibinfo {author} {\bibfnamefont {N.}~\bibnamefont {Werthamer}}, \bibinfo {author} {\bibfnamefont {E.}~\bibnamefont {Helfand}},\ and\ \bibinfo {author} {\bibfnamefont {P.}~\bibnamefont {Hohenberg}},\ }\bibfield  {title} {\bibinfo {title} {Temperature and purity dependence of the superconducting critical field, h c 2. iii. electron spin and spin-orbit effects},\ }\href {https://doi.org/10.1103/PhysRev.147.295} {\bibfield  {journal} {\bibinfo  {journal} {Physical Review}\ }\textbf {\bibinfo {volume} {147}},\ \bibinfo {pages} {295} (\bibinfo {year} {1966})}\BibitemShut {NoStop}%
\bibitem [{\citenamefont {Gorter}\ and\ \citenamefont {Casimir}(1934)}]{gorter1934supraconductivity}%
  \BibitemOpen
  \bibfield  {author} {\bibinfo {author} {\bibfnamefont {C.~J.}\ \bibnamefont {Gorter}}\ and\ \bibinfo {author} {\bibfnamefont {H.}~\bibnamefont {Casimir}},\ }\bibfield  {title} {\bibinfo {title} {On supraconductivity i},\ }\href {https://doi.org/10.1016/S0031-8914(34)90037-9} {\bibfield  {journal} {\bibinfo  {journal} {Physica}\ }\textbf {\bibinfo {volume} {1}},\ \bibinfo {pages} {306} (\bibinfo {year} {1934})}\BibitemShut {NoStop}%
\bibitem [{\citenamefont {Rata}\ and\ \citenamefont {Hibma}(2005)}]{rata2005strain}%
  \BibitemOpen
  \bibfield  {author} {\bibinfo {author} {\bibfnamefont {A.~D.}\ \bibnamefont {Rata}}\ and\ \bibinfo {author} {\bibfnamefont {T.}~\bibnamefont {Hibma}},\ }\bibfield  {title} {\bibinfo {title} {Strain-induced properties of epitaxial vox thin films},\ }\href {https://doi.org/10.1140/epjb/e2005-00042-6} {\bibfield  {journal} {\bibinfo  {journal} {The European Physical Journal B-Condensed Matter and Complex Systems}\ }\textbf {\bibinfo {volume} {43}},\ \bibinfo {pages} {195} (\bibinfo {year} {2005})}\BibitemShut {NoStop}%
\bibitem [{\citenamefont {Wander}\ \emph {et~al.}(2003)\citenamefont {Wander}, \citenamefont {Bush},\ and\ \citenamefont {Harrison}}]{wander2003stability}%
  \BibitemOpen
  \bibfield  {author} {\bibinfo {author} {\bibfnamefont {A.}~\bibnamefont {Wander}}, \bibinfo {author} {\bibfnamefont {I.}~\bibnamefont {Bush}},\ and\ \bibinfo {author} {\bibfnamefont {N.}~\bibnamefont {Harrison}},\ }\bibfield  {title} {\bibinfo {title} {Stability of rocksalt polar surfaces: An ab initio study of mgo (111) and nio (111)},\ }\href {https://doi.org/10.1103/PhysRevB.68.233405} {\bibfield  {journal} {\bibinfo  {journal} {Physical Review B}\ }\textbf {\bibinfo {volume} {68}},\ \bibinfo {pages} {233405} (\bibinfo {year} {2003})}\BibitemShut {NoStop}%
\bibitem [{\citenamefont {Liu}\ \emph {et~al.}(2006)\citenamefont {Liu}, \citenamefont {Liu}, \citenamefont {Zheng},\ and\ \citenamefont {Jiang}}]{liu2006surface}%
  \BibitemOpen
  \bibfield  {author} {\bibinfo {author} {\bibfnamefont {W.}~\bibnamefont {Liu}}, \bibinfo {author} {\bibfnamefont {X.}~\bibnamefont {Liu}}, \bibinfo {author} {\bibfnamefont {W.}~\bibnamefont {Zheng}},\ and\ \bibinfo {author} {\bibfnamefont {Q.}~\bibnamefont {Jiang}},\ }\bibfield  {title} {\bibinfo {title} {Surface energies of several ceramics with nacl structure},\ }\href {https://doi.org/10.1016/j.susc.2005.10.035} {\bibfield  {journal} {\bibinfo  {journal} {Surface science}\ }\textbf {\bibinfo {volume} {600}},\ \bibinfo {pages} {257} (\bibinfo {year} {2006})}\BibitemShut {NoStop}%
\bibitem [{\citenamefont {Tellekamp}\ \emph {et~al.}(2017)\citenamefont {Tellekamp}, \citenamefont {Shank},\ and\ \citenamefont {Doolittle}}]{tellekamp2017molecular}%
  \BibitemOpen
  \bibfield  {author} {\bibinfo {author} {\bibfnamefont {M.~B.}\ \bibnamefont {Tellekamp}}, \bibinfo {author} {\bibfnamefont {J.~C.}\ \bibnamefont {Shank}},\ and\ \bibinfo {author} {\bibfnamefont {W.~A.}\ \bibnamefont {Doolittle}},\ }\bibfield  {title} {\bibinfo {title} {Molecular beam epitaxy of lithium niobium oxide multifunctional materials},\ }\href {https://doi.org/10.1016/j.jcrysgro.2017.02.020} {\bibfield  {journal} {\bibinfo  {journal} {Journal of Crystal Growth}\ }\textbf {\bibinfo {volume} {463}},\ \bibinfo {pages} {156} (\bibinfo {year} {2017})}\BibitemShut {NoStop}%
\bibitem [{\citenamefont {Bolzan}\ \emph {et~al.}(1994)\citenamefont {Bolzan}, \citenamefont {Fong}, \citenamefont {Kennedy},\ and\ \citenamefont {Howard}}]{bolzan1994powder}%
  \BibitemOpen
  \bibfield  {author} {\bibinfo {author} {\bibfnamefont {A.~A.}\ \bibnamefont {Bolzan}}, \bibinfo {author} {\bibfnamefont {C.}~\bibnamefont {Fong}}, \bibinfo {author} {\bibfnamefont {B.~J.}\ \bibnamefont {Kennedy}},\ and\ \bibinfo {author} {\bibfnamefont {C.~J.}\ \bibnamefont {Howard}},\ }\bibfield  {title} {\bibinfo {title} {A powder neutron diffraction study of semiconducting and metallic niobium dioxide},\ }\href {https://doi.org/10.1006/jssc.1994.1334} {\bibfield  {journal} {\bibinfo  {journal} {Journal of Solid State Chemistry}\ }\textbf {\bibinfo {volume} {113}},\ \bibinfo {pages} {9} (\bibinfo {year} {1994})}\BibitemShut {NoStop}%
\bibitem [{\citenamefont {Nair}\ \emph {et~al.}(2018)\citenamefont {Nair}, \citenamefont {Ruf}, \citenamefont {Schreiber}, \citenamefont {Miao}, \citenamefont {Grandon}, \citenamefont {Baek}, \citenamefont {Goodge}, \citenamefont {Ruff}, \citenamefont {Kourkoutis}, \citenamefont {Shen} \emph {et~al.}}]{nair2018demystifying}%
  \BibitemOpen
  \bibfield  {author} {\bibinfo {author} {\bibfnamefont {H.~P.}\ \bibnamefont {Nair}}, \bibinfo {author} {\bibfnamefont {J.~P.}\ \bibnamefont {Ruf}}, \bibinfo {author} {\bibfnamefont {N.~J.}\ \bibnamefont {Schreiber}}, \bibinfo {author} {\bibfnamefont {L.}~\bibnamefont {Miao}}, \bibinfo {author} {\bibfnamefont {M.~L.}\ \bibnamefont {Grandon}}, \bibinfo {author} {\bibfnamefont {D.~J.}\ \bibnamefont {Baek}}, \bibinfo {author} {\bibfnamefont {B.~H.}\ \bibnamefont {Goodge}}, \bibinfo {author} {\bibfnamefont {J.~P.}\ \bibnamefont {Ruff}}, \bibinfo {author} {\bibfnamefont {L.~F.}\ \bibnamefont {Kourkoutis}}, \bibinfo {author} {\bibfnamefont {K.~M.}\ \bibnamefont {Shen}}, \emph {et~al.},\ }\bibfield  {title} {\bibinfo {title} {Demystifying the growth of superconducting sr2ruo4 thin films},\ }\bibfield  {journal} {\bibinfo  {journal} {APL Materials}\ }\textbf {\bibinfo {volume} {6}},\ \href {https://doi.org/10.1063/1.5053084} {10.1063/1.5053084} (\bibinfo {year} {2018})\BibitemShut {NoStop}%
\bibitem [{\citenamefont {Ahadi}\ \emph {et~al.}(2019)\citenamefont {Ahadi}, \citenamefont {Galletti}, \citenamefont {Li}, \citenamefont {Salmani-Rezaie}, \citenamefont {Wu},\ and\ \citenamefont {Stemmer}}]{ahadi2019enhancing}%
  \BibitemOpen
  \bibfield  {author} {\bibinfo {author} {\bibfnamefont {K.}~\bibnamefont {Ahadi}}, \bibinfo {author} {\bibfnamefont {L.}~\bibnamefont {Galletti}}, \bibinfo {author} {\bibfnamefont {Y.}~\bibnamefont {Li}}, \bibinfo {author} {\bibfnamefont {S.}~\bibnamefont {Salmani-Rezaie}}, \bibinfo {author} {\bibfnamefont {W.}~\bibnamefont {Wu}},\ and\ \bibinfo {author} {\bibfnamefont {S.}~\bibnamefont {Stemmer}},\ }\bibfield  {title} {\bibinfo {title} {Enhancing superconductivity in srtio3 films with strain},\ }\href {https://doi.org/10.1126/sciadv.aaw0120} {\bibfield  {journal} {\bibinfo  {journal} {Science advances}\ }\textbf {\bibinfo {volume} {5}},\ \bibinfo {pages} {eaaw0120} (\bibinfo {year} {2019})}\BibitemShut {NoStop}%
\bibitem [{\citenamefont {Hubbard}\ and\ \citenamefont {Schlom}(1996)}]{hubbard1996thermodynamic}%
  \BibitemOpen
  \bibfield  {author} {\bibinfo {author} {\bibfnamefont {K.}~\bibnamefont {Hubbard}}\ and\ \bibinfo {author} {\bibfnamefont {D.}~\bibnamefont {Schlom}},\ }\bibfield  {title} {\bibinfo {title} {Thermodynamic stability of binary oxides in contact with silicon},\ }\href {https://doi.org/10.1557/JMR.1996.0350} {\bibfield  {journal} {\bibinfo  {journal} {Journal of Materials Research}\ }\textbf {\bibinfo {volume} {11}},\ \bibinfo {pages} {2757} (\bibinfo {year} {1996})}\BibitemShut {NoStop}%
\bibitem [{\citenamefont {McRae}\ \emph {et~al.}(2020)\citenamefont {McRae}, \citenamefont {Wang}, \citenamefont {Gao}, \citenamefont {Vissers}, \citenamefont {Brecht}, \citenamefont {Dunsworth}, \citenamefont {Pappas},\ and\ \citenamefont {Mutus}}]{mcrae2020materials}%
  \BibitemOpen
  \bibfield  {author} {\bibinfo {author} {\bibfnamefont {C.~R.~H.}\ \bibnamefont {McRae}}, \bibinfo {author} {\bibfnamefont {H.}~\bibnamefont {Wang}}, \bibinfo {author} {\bibfnamefont {J.}~\bibnamefont {Gao}}, \bibinfo {author} {\bibfnamefont {M.~R.}\ \bibnamefont {Vissers}}, \bibinfo {author} {\bibfnamefont {T.}~\bibnamefont {Brecht}}, \bibinfo {author} {\bibfnamefont {A.}~\bibnamefont {Dunsworth}}, \bibinfo {author} {\bibfnamefont {D.~P.}\ \bibnamefont {Pappas}},\ and\ \bibinfo {author} {\bibfnamefont {J.}~\bibnamefont {Mutus}},\ }\bibfield  {title} {\bibinfo {title} {Materials loss measurements using superconducting microwave resonators},\ }\bibfield  {journal} {\bibinfo  {journal} {Review of Scientific Instruments}\ }\textbf {\bibinfo {volume} {91}},\ \href {https://doi.org/10.1063/5.0017378} {10.1063/5.0017378} (\bibinfo {year} {2020})\BibitemShut {NoStop}%
\bibitem [{\citenamefont {Siddiqi}(2021)}]{siddiqi2021engineering}%
  \BibitemOpen
  \bibfield  {author} {\bibinfo {author} {\bibfnamefont {I.}~\bibnamefont {Siddiqi}},\ }\bibfield  {title} {\bibinfo {title} {Engineering high-coherence superconducting qubits},\ }\href {https://doi.org/10.1038/s41578-021-00370-4} {\bibfield  {journal} {\bibinfo  {journal} {Nature Reviews Materials}\ }\textbf {\bibinfo {volume} {6}},\ \bibinfo {pages} {875} (\bibinfo {year} {2021})}\BibitemShut {NoStop}%
\end{thebibliography}
%

\renewcommand{\thefigure}{S\arabic{figure}}

\setcounter{figure}{0}

\begin{table*}[h]
\centering
\caption{
	\label{table} 
	Structural data and epitaxial relationships
      }
        \begin{tabular}{c|c|cc}\hline\hline
          Compounds & Space group &  ~~~~~Out-of-plane~~~~~ & ~~~~~In-plane~~~~~ \\\hline
          \AO & $R\overline{3}c$ (167) & [0001] & [11$\overline{2}$0]\\\hline
          Nb & $Im\overline{3}m$ (229) & [110]  &  [1$\overline{1}$0]\\
          NbO & $Pm\overline{3}m$ (221) &  [111] & [11$\overline{2}$]\\
          NbO$_2$ & $P4$$_2$$/mnm$ (136)  &  [100] & [010]\\
    \hline\hline
  \end{tabular}
\end{table*}

\begin{figure*}[t]
  \centering
   \includegraphics[width=170mm]{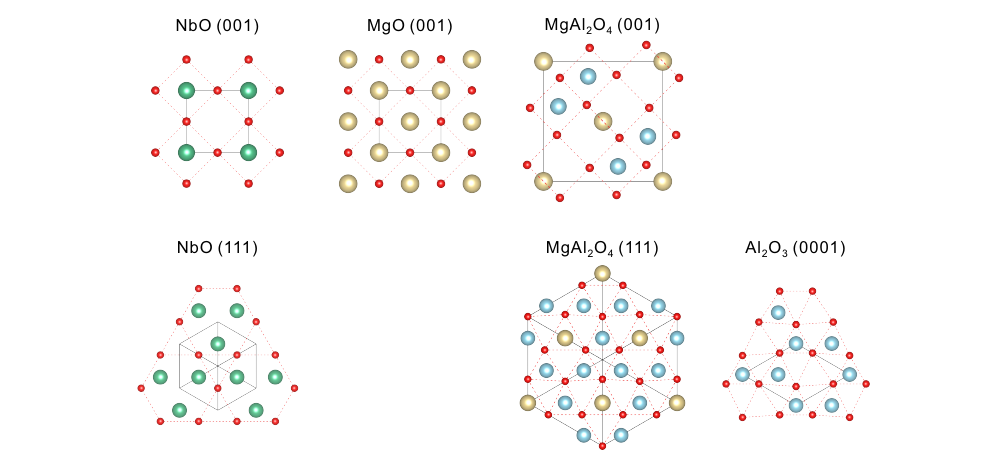}
   \caption{Atomic structures of effective oxygen lattices of vacancy-ordered rock-salt NbO, rock-salt MgO, spinel \MAO, and corundum \AO.
   }
   \label{FigS1}
\end{figure*}

\begin{figure*}[t]
  \centering
   \includegraphics[width=170mm]{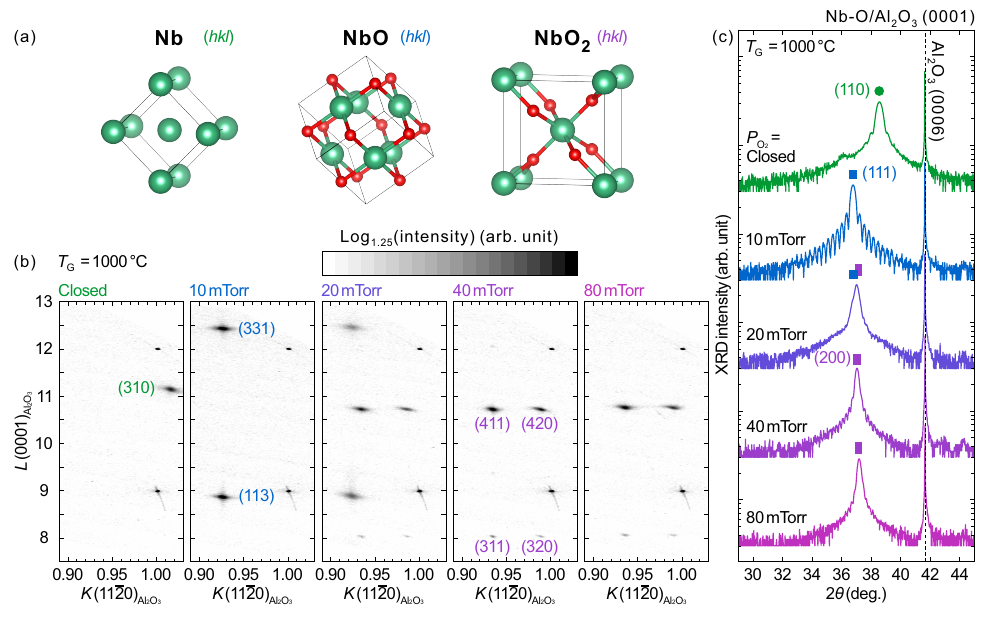}
   \caption{XRD analysis of Nb-O on~\AO~(0001). (a) Crystal structures of body-centered cubic Nb, vacancy-ordered rock-salt NbO, and rutile NbO$_2$. (b,c) XRD reciprocal space mapping (b) and 2$\theta$-$\omega$ (c) measurements of the Nb-O/\AO~(0001) samples grown at~\Tg~=~1000~\dC~and~\Po~= `Closed', 10~mTorr, 20~mTorr, 40~mTorr, and 80~mTorr.
   }
   \label{FigS2}
\end{figure*}

\begin{figure*}[t]
  \centering
   \includegraphics[width=170mm]{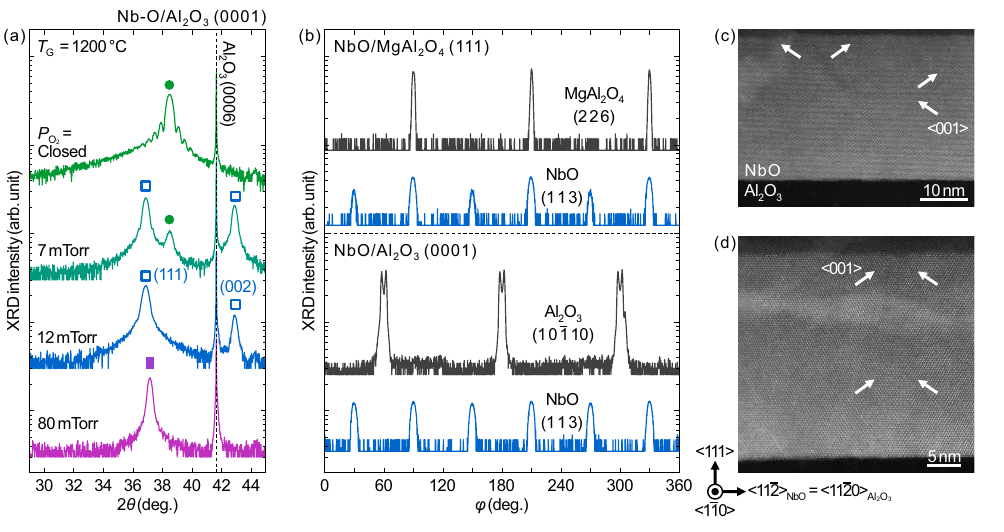}
   \caption{Structural imperfections of epitaxial NbO (111) films. (a) Out-of-plane XRD of the Nb-O/\AO~(0001) samples grown at~\Tg~=~1200~\dC~and~\Po~= `Closed', 7~mTorr, 12~mTorr, and 80~mTorr. NbO (111) films tend to grow polycrystalline at~\Tg~=~1200~\dC. (b) XRD phi scan of NbO/\MAO~(111) and NbO\AO~(0001) samples grown at~\Tg~=~1000~\dC~and~\Po~= 10~mTorr. Both samples exhibit six-fold symmetry of the NbO (113) peak, indicating twin domain formation. (c,d) HAADF-STEM images of twin domain boundaries in the NbO\AO~(0001) sample taken from the~\AO~[1$\overline{1}$00] zone axis. 
   }
   \label{FigS3}
\end{figure*}

\end{document}